\documentclass[journal]{IEEEtran}
\usepackage{graphicx}
\usepackage{graphics}

\begin{document}
\title{R\&D of Commercially Manufactured Large GEM Foils}

\author{M.~Posik and~B.~Surrow\thanks{Manuscript received November 23, 2015. This work was supported in part by an EIC grant, subtracted by Brookhaven Science Associates \#223228.}\thanks{M.~Posik is with Temple University Philadelphia, PA 19122 USA (e-mail:posik@temple.edu).}\thanks{B.~Surrow is with Temple University Philadelphia, PA 19122 USA (e-mail: surrow@temple.edu).}}
%%        and~B.~Surrow% <-this % stops a space
%%%        and~Third~C.Author,~\IEEEmembership{Life~Fellow,~IEEE}% <-this % stops a space
%%%\thanks{Manuscript received November 23, 2015. This work was supported in part by an EIC grant, subtracted by Brookhaven Science Associates #223228.}% <-this % stops a space
%%\thanks{M.~Posik is with Temple University Philadelphia, PA 19122 USA (e-mail: %%posik@temple.edu).}%
%%\thanks{B.~Surrow is with Temple University Philadelphia, PA 19122 USA (e-mail: %%surrow@temple.edu).}%
%%}

\maketitle
\pagestyle{empty}
\thispagestyle{empty}

\begin{abstract}
Many experiments are currently using or proposing to use large area GEM foils in their detectors, which is creating a need for commercially available GEM foils. Currently CERN is the only main distributor of GEM foils, however with the growing interest in GEM technology keeping up with the increasing demand for GEM foils will be difficult. Thus the commercialization of GEM foils has been established by Tech-Etch Inc. of Plymouth, MA, USA using the single-mask technique, which is capable of producing GEM foils over a meter long.

To date Tech-Etch has successfully manufactured 10 $\times$ 10 cm$^2$ and 40 $\times$ 40 cm$^2$ GEM foils. We will report on the electrical and geometrical properties, along with the inner and outer hole diameter size uniformity of these foils. Furthermore, Tech-Etch has now begun producing even larger GEM foils of 50 $\times$ 50 cm$^2$, and are currently looking into how to accommodate GEM foils on the order of one meter long. 

The Tech-Etch foils were found to have excellent electrical properties. 
The measured mean optical properties were found to reflect the desired parameters and are consistent with those measured in double-mask GEM foils, as well as single-mask GEM foils produced at CERN. They also show good hole diameter uniformity over the active area.\end{abstract}

%\begin{IEEEkeywords}
%IEEEtran, journal, \LaTeX, paper, template.
%\end{IEEEkeywords}

\section{Introduction}
\IEEEPARstart{G}{as} electron multipliers (GEMs) were invented in 1997~\cite{Sauli:1997qp}, and since that time have been making their presence known in the nuclear and particle physics communities by incorporating them into various detectors~\cite{Surrow:2010zza,Altunbas:2002ds}. With the successful use of GEM technology, many future experiments and experiment upgrades are either planning on or looking into using GEM foils, such as ALICE~\cite{Gasik:2014sga}, JLab's Super BigBite Spectrometer~\cite{SBS}, CMS~\cite{Abbaneo:2014} and a potential electron ion collider (EIC)~\cite{EIC}. 

The main distributor of GEM foils to the scientific community is CERN. In the past CERN has been able to adequately provide GEM foils to experiments that needed them. However, given the newly generated interest in GEM foils and the fact that CERN is not a dedicated production facility, one can not expect CERN to be able to provide all experiments with the GEM foils that they need. As a result CERN has been working with Tech-Etch Inc~\cite{TechEtch} to transfer its technology in efforts to commercialize GEM foils. Tech-Etch Inc. is a company based in Plymouth, Massachusetts who have produced large area (up to $\sim$50$\times$50 cm$^{2}$) GEM foils~\cite{Surrow:2010zza,Becker:2006,Surrow:2007,Simon:2007,Simon:2009,posik:2015nim} using CERN's patented single-mask and double-mask etching processes~\cite{Villa:2010w}. The single-mask etching process allows for GEM foils to be on the order of one meter long, which is crucial for future use as experiments will require larger coverage.          

\section{Tech-Etch GEM Production}
Tech-Etch has successfully produced 10$\times$10, 40$\times$40, and 50$\times$50 cm$^2$ single-mask GEM foils, which have been sent to Temple University for analysis of their electrical performance and geometrical properties. A complete analysis of the optical and electrical performance of the 10$\times$10 and 40$\times$40 cm$^2$ GEM foils is discussed in reference~\cite{posik:2015nim}. A summary of our measurement techniques and results will be presented in Sections~\ref{sec:measure} and~\ref{sec:results}, respectively. The commercialization of the 50$\times$50 cm$^2$ GEM foils is still ongoing as Tech-Etch continues to tweak their production parameters in order to optimize the GEM foil quality. However given the initial success of the 50$\times$50 cm$^2$ GEM foils, Tech-Etch is currently looking into upgrading their facilities in order to accommodate GEM foils that are on the order of one meter long. These meter long foils would ultimately be used in an EIC prototype tracking detector briefly touched on in Section~\ref{sec:EICGEM}, with more detailed information found in reference~\cite{Zhang}.   

\section{Measurement Techniques}\label{sec:measure}
The production quality of a GEM foil can be quantified through its electrical and geometrical properties. The electrical properties of the GEM foil were determined by measuring its leakage current and charge accumulation over an extended period of time. The leakage current was measured by applying a high voltage across the foil and measuring the resulting current. The charge accumulation was measured by applying a fixed high voltage across the foil and measuring the resulting current as a function of time. 
The geometrical quality is determined through an optical analysis using an automated 2D CCD scanner, which is capable of scanning the entire active area of the GEM foil. Several quantities where measured, such as the pitch between two neighboring holes, the inner hole diameter (determined from the polyimide (Apical) layer) and the outer hole diameter (determined from the copper layer).

\subsection{Electrical Analysis}
The leakage current and charge accumulation measurements were carried out in a class 1,000 clean room. The GEM foils were placed in a plexiglass enclosure which was flushed with nitrogen. After about an hour of flushing time, the leakage current was measured at 100 V increments from 0 to 600 V. The charge accumulation was measured by applying a fixed voltage of 550 V to the GEM foil and measuring its leakage current periodically over the course of a few days.   

\subsection{Optical Analysis} 
The optical properties of the GEM foils were measured with the use of an automated 2D CCD scanner. The setup used at Temple University is identical to that which is described in reference~\cite{Becker:2006}. The CCD camera setup consists of a video camera connected to a 12x zoom lens through a 2x adapter, with a ring of LEDs around the lens face (front light). The CCD setup is coupled to a support stage with a LED light mounted below it (back light). The stage is able to traverse in two dimensions, covering an area of about 30$\times$15 cm$^2$. The apparatus is controlled through a MATLAB~\cite{Matlab} graphical interface. The sensitivity of the CCD camera to the GEM foil's inner or outer hole diameters is determined by the lighting scheme used. If the front light is used to illuminate the GEM foil, then we are sensitive to the GEM's outer hole diameters. On the other hand, if the back light is used to illuminate the GEM foil, then the measurements will be sensitive to the GEM foil's inner hole diameters. By using MATLAB to analyze the images and convert pixel counts into distances, the pitch and the inner and outer hole diameters can be determined. Due to the limited stage size, the larger GEM foils ({\it{i.e.}} the 40$\times$40 cm$^2$) need to be divided into several CCD scan regions in order to image the entire active area of a GEM foil.

\section{Results}\label{sec:results}

\subsection{Electrical Performance}
The leakage currents of all Tech-Etch single-mask GEM foils were independently measured twice, once by Tech-Etch prior to shipping the foils to Temple University and again by Temple University after shipping. Measurements were consistent between Tech-Etch and Temple University and showed a typical leakage current to be around 1 nA on all foils. The leakage current from several CERN 10$\times$10 cm$^2$ foils were also measured to provide a direct comparison to the Tech-Etch electrical performance. The CERN foils were found to also display typical leakage currents of around 1 nA. A random selection of Tech-Etch single-mask foils were selected and had 550 V applied to them. No significant charge accumulation was observed while monitoring the leakage current over the course of several days.  

\subsection{Optical Quality}
After optically scanning and measuring all of the Tech-Etch single-mask GEM foils, it was found that there was good inner and outer hole diameter uniformity across the GEM foil, and that the mean values of the pitch, inner and outer hole diameters are in line with the values measured previously from double-mask GEM foils as well as single-mask GEM foils produced by CERN. 

The pitch of the Tech-Etch GEM foils was found to be consistent for each foil, including between the 10$\times$10 and 40$\times$40 cm$^2$ foils, at about 138$\mu m$. 

\begin{figure}[!t]
\centering
\includegraphics[width=\columnwidth]{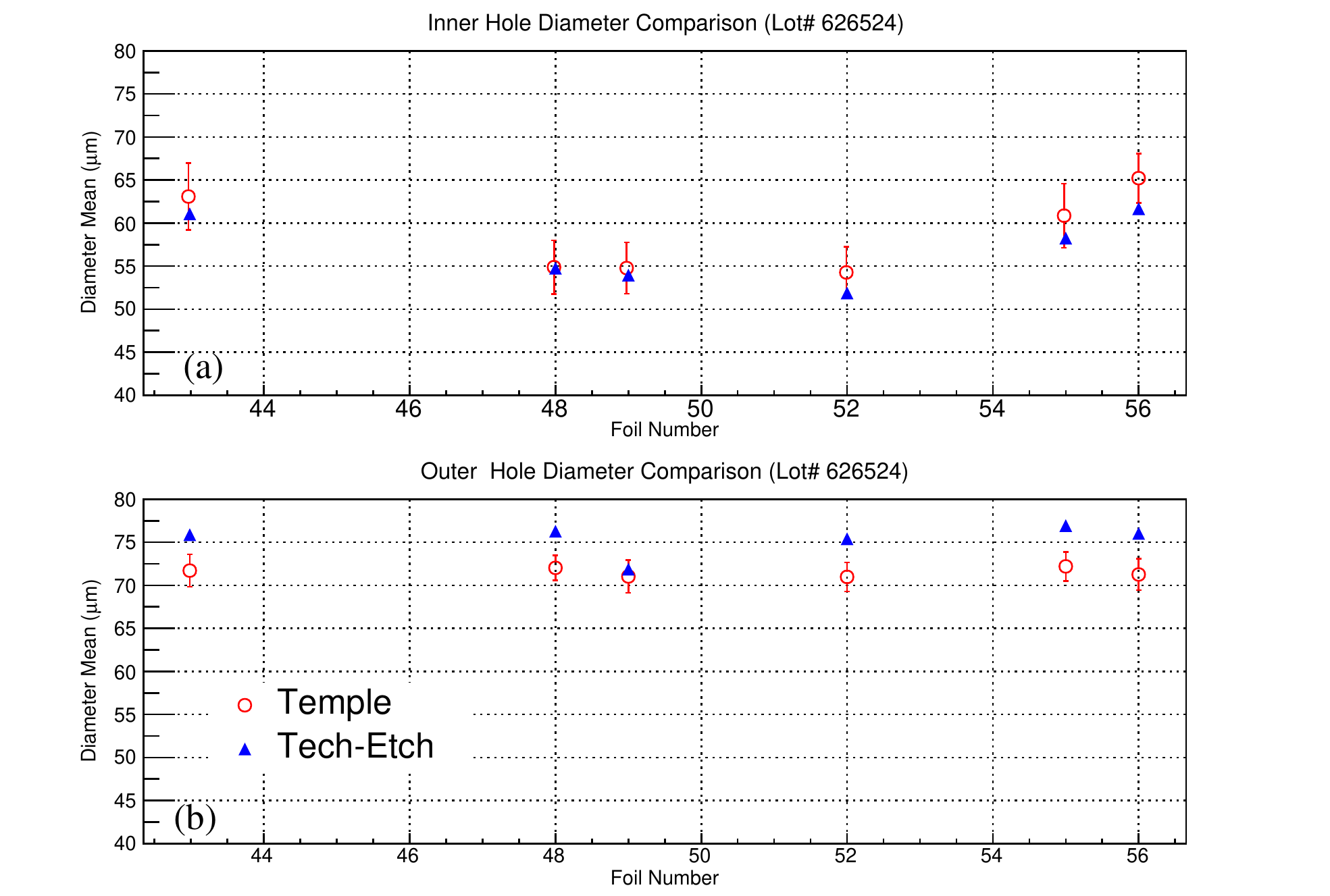}
%\includegraphics[width=3.5in]{CommonFoil.png}
% where an .eps filename suffix will be assumed under latex, 
% and a .pdf suffix will be assumed for pdflatex; or what has been declared
% via \DeclareGraphicsExtensions.
\caption{Inner (a) and outer (b) hole diameter measurements done by Tech-Etch and Temple University on 10$\times$10 cm$^2$ single-mask GEM foils produced by Tech-Etch.}
\label{fig:10by10Comp}
\end{figure}

The inner and outer hole diameters for the 10$\times$10 cm$^2$ single-mask Tech-Etch GEM fois are shown in Fig.~\ref{fig:10by10Comp}, where the vertical axis lists the mean inner (upper panel) and outer (lower panel) diameters per foil, and the horizontal axis corresponds to the foil that was measured. The solid triangle markers show measurement made by Tech-Etch, where they randomly considered only about 9 holes measurements from high resolution images over the entire active area of a GEM foil. There are no error bars calculated for the Tech-Etch measurements. The open circle markers show the measurements done by Temple University where the entire active area of a GEM foil was considered. The error bar shown on these points represent the spread of the respective diameter distribution.

\begin{figure}[!t]
\centering
\includegraphics[width=\columnwidth]{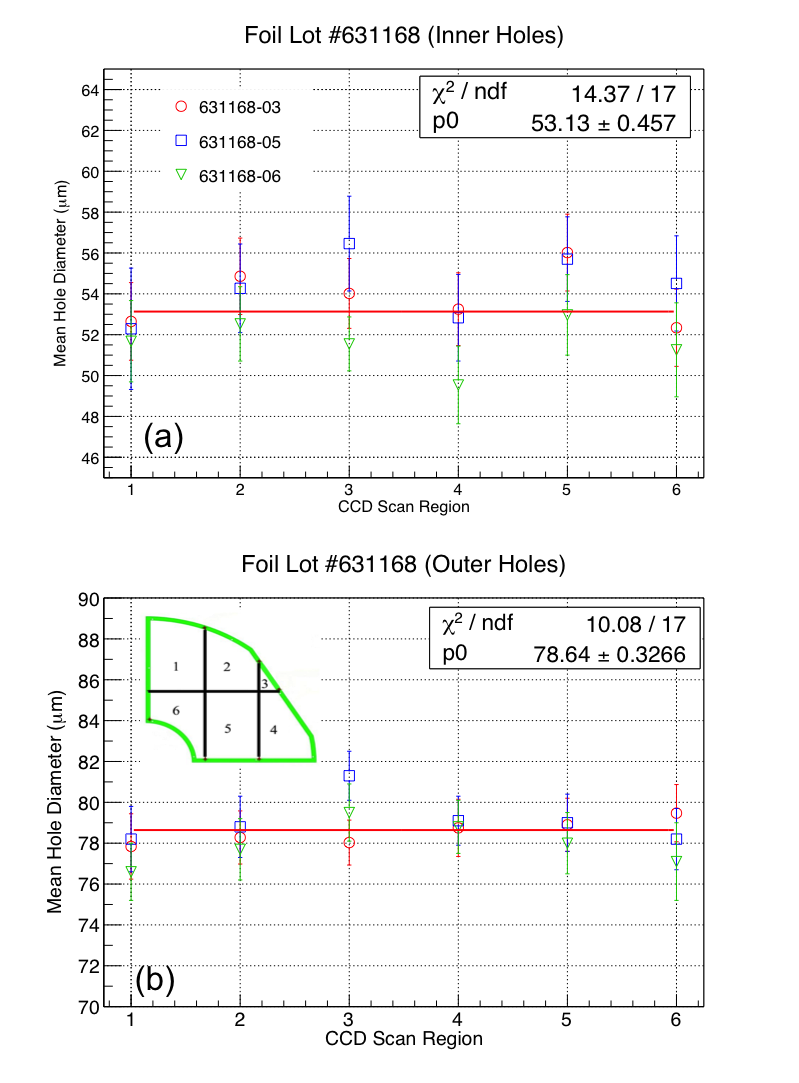}
%\includegraphics[width=3.5in]{CommonFoil.png}
% where an .eps filename suffix will be assumed under latex, 
% and a .pdf suffix will be assumed for pdflatex; or what has been declared
% via \DeclareGraphicsExtensions.
\caption{Inner (a) and outer (b) hole diameter measurements done by Temple University on 40$\times$40 cm$^2$ single-mask GEM foils produced by Tech-Etch.}
\label{fig:40by40Comp}
\end{figure}

The same analysis was also done for the Tech-Etch single mask 40$\times$40 cm$^2$ foils with the results summarized in Fig.~\ref{fig:40by40Comp}. The division into 6 CCD scan regions of the active area of the GEM foil is shown in the top left corner in the bottom panel. The top (bottom) panel shows the mean inner (outer) hole diameter for each CCD scan region, where each marker type represents a different 40$\times$40 cm$^2$ GEM foil. Again the error bars in these plots represent the spread in the diameter distribution for that given foil at that specific CCD scan region. These data were then fit with at constant line to obtain a mean diameter value across all of the 40$\times$40 cm$^2$ GEM foils. From this figure one can see very consistent hole diameters across all of the foils. 

\begin{figure}[!t]
\centering
\includegraphics[width=\columnwidth]{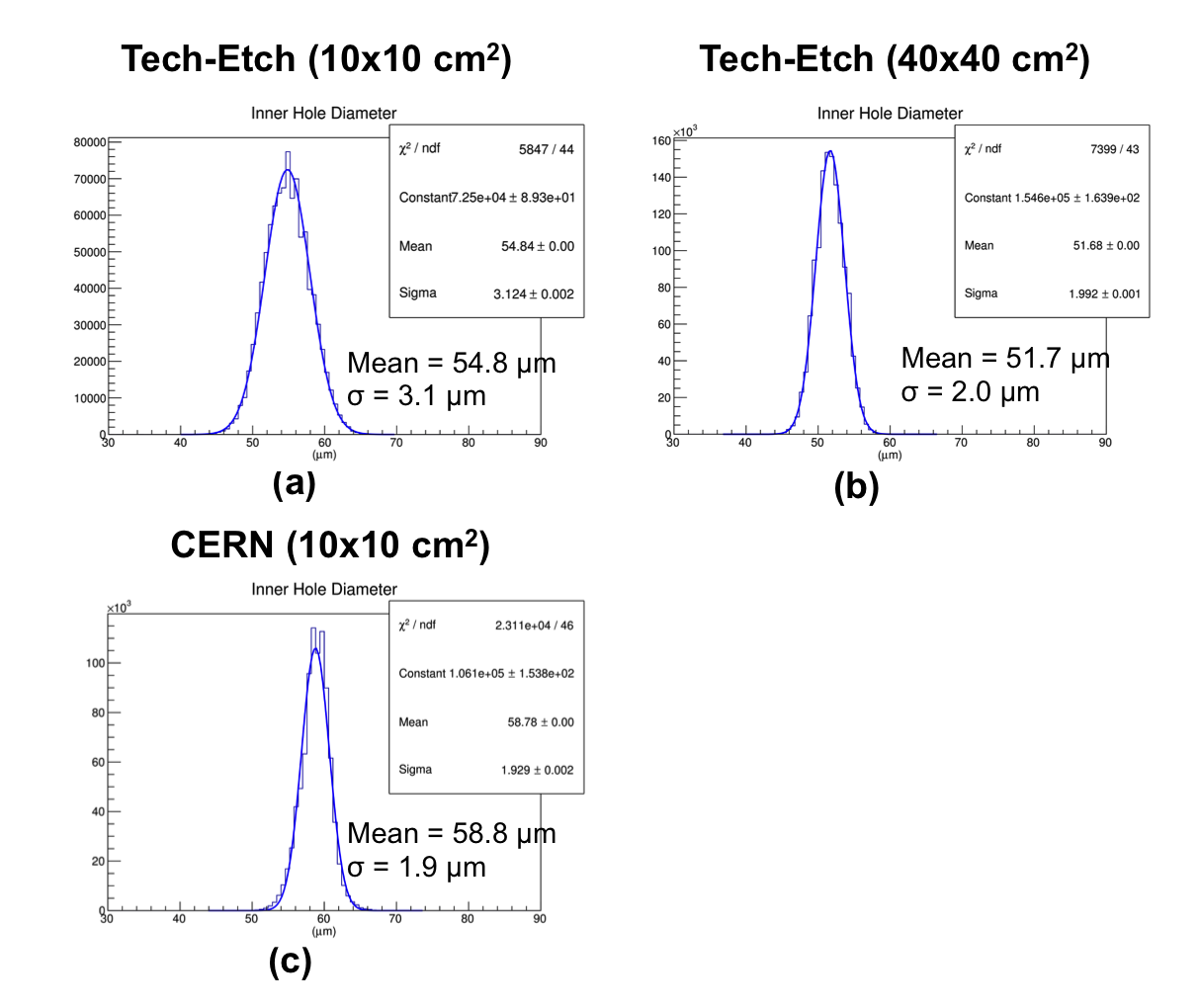}
%\includegraphics[width=3.5in]{CommonFoil.png}
% where an .eps filename suffix will be assumed under latex, 
% and a .pdf suffix will be assumed for pdflatex; or what has been declared
% via \DeclareGraphicsExtensions.
\caption{Inner hole diameter distributions from selected single-mask GEM foils. (a) a Tech-Etch 10$\times$10 cm$^2$ GEM foil, (b) a Tech-Etch 40$\times$40 cm$^2$ GEM foil, and (c) a CERN 10$\times$10 cm$^2$ GEM foil.}
\label{fig:InDist}
\end{figure}

Like with the electrical performance, several CERN produced 10$\times$10 cm$^2$ single-mask GEM foils were optically measured to provide a source of reference for the Tech-Etch foils. The CERN measurements yielded similar results for the mean hole diameter sizes. While the outer hole diameter distributions were similar between the Tech-Etch and CERN foils, the CERN foils displayed slightly more uniform inner hole diameters than the Tech-Etch single-mask 10$\times$10 cm$^2$, which when fitted with a Gaussian distribution had a sigma about 1 $\mu m$ narrower. However the inner diameter distribution of the Tech-Etch 40$\times$40 cm$^2$ single-mask GEM foils showed a similar uniformity as the CERN 10$\times$10 cm$^2$ GEM foils. A representative comparision between the three foil types can be seen in Fig.~\ref{fig:InDist}.      

\section{Common GEM Foil}\label{sec:EICGEM}
\begin{figure}[!t]
\centering
\includegraphics[scale=0.5]{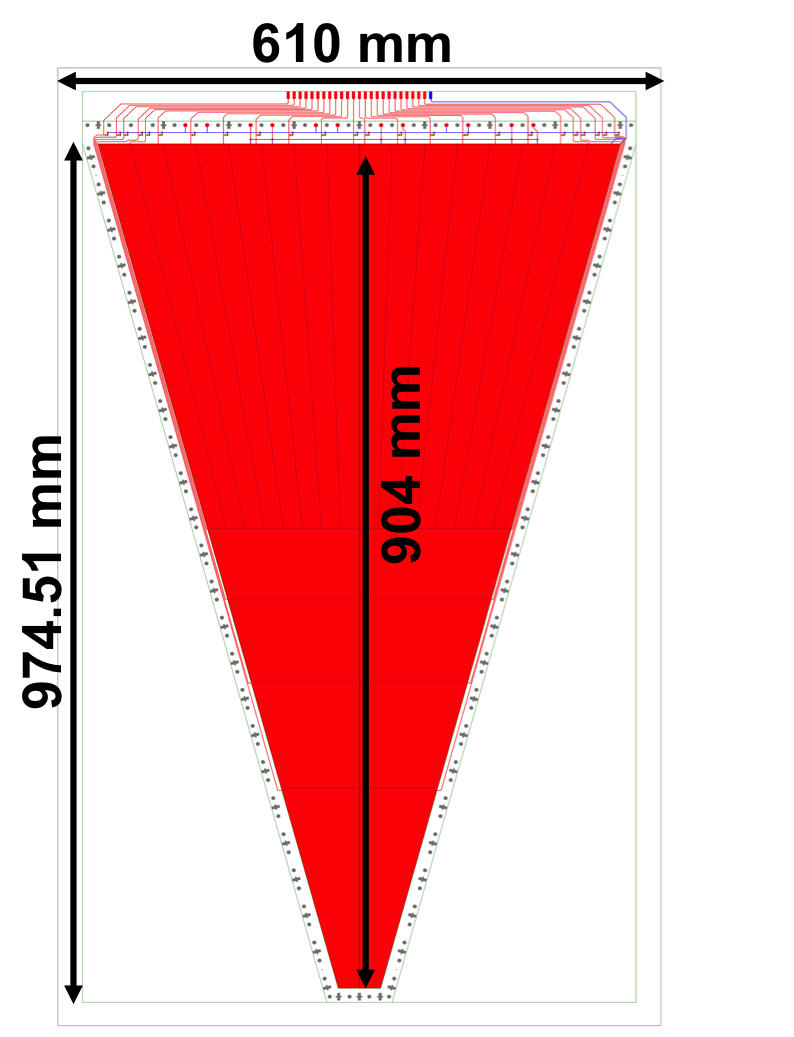}
%\includegraphics[width=\columnwidth]{CommonFoil.png}
%\includegraphics[width=3.5in]{CommonFoil.png}
% where an .eps filename suffix will be assumed under latex, 
% and a .pdf suffix will be assumed for pdflatex; or what has been declared
% via \DeclareGraphicsExtensions.
\caption{Design of common GEM foil to be used in EIC prototype tracking detectors. Image produced by A. Zhang.}
\label{fig:commonfoil}
\end{figure}

Several new tracking prototype detectors are being designed and built for potential use in an EIC if it were to be built. These tracking detectors would make use of large area GEM foils ($\sim$ 1 m long). Through a collaborative effort between Florida Institute of Technology (FIT), University of Virgina (UVa), and Temple University (TU) a GEM foil design was agreed upon. The foil is shown in Fig.~\ref{fig:commonfoil} and has a foil area of about 97$\times$60 cm$^2$. The foil has an opening angle of 30.1$^\circ$ and is split into 24 HV segments with each having an area of about 107 cm$^2$. Of the 24 HV segments 16 of them are azimuthal and 8 are radial HV segments. The total tracking detector will use 12 of the trapezoidal triple-GEM detectors arranged in a disk to achieve the desired acceptance. This foil design will be used by each institute FIT, UVa, and TU to design three different tracking detectors. The foil design and production has already been discussed with CERN, and it is our intention to also procure common GEM foils from Tech-Etch once their facilities are able to accommodate the larger foil size. More details on the common GEM foil design and the various tracking detector prototypes can be found in reference~\cite{Zhang}.

\section*{Acknowledgment}
We would like to thank David Crary, Kerry Kearney, and Matthew Campbell (Tech-Etch Inc.), as well as M.~Hohlmann (FIT), R.~Majka (Yale), and especially R.~De~Oliveira (CERN) for their useful discussions, guidance, and expertise which has lead to the successful commercialization of GEM technology.


%\bibliographystyle{IEEEtran}

\end{document}